\documentclass{jltp}
\usepackage{amsfonts}
\usepackage{amsmath}
\usepackage{amssymb}
\usepackage[final,dvips]{graphicx}

\title{Tilted and crossing vortex chains in layered
superconductors}

\author{A. E. Koshelev}

\address{Materials Science Division, Argonne National Laboratory,Argonne, Illinois 60439}

\runninghead{A. E. Koshelev}{Tilted and Crossing Vortex Chains in
Layered Superconductors}

\begin{document}

\maketitle

\begin{abstract}
In the presence of the Josephson vortex lattice in layered
superconductors, a small c-axis magnetic field penetrates in the
form of vortex chains.  In general, the structure of a single
chain is determined by the ratio of the London [$\lambda$] and
Josephson [$\lambda_{J}$] lengths, $\alpha= \lambda/\lambda_{J}$.
The chain is composed of tilted vortices at large $\alpha$'s
(tilted chain) and at small $\alpha$'s it consists of a crossing
array of Josephson vortices and pancake-vortex stacks (crossing
chain). We study chain structures at intermediate $\alpha$'s and
found two types of phase transitions. For $\alpha\lesssim 0.6$ the
ground state is given by the crossing chain in a wide range of
pancake separations $a\gtrsim [2-3]\lambda_J$. However, due to
attractive coupling between deformed pancake stacks, the
equilibrium separation can not exceed some maximum value depending
on the in-plane field and $\alpha$. The first phase transition
takes place with decreasing pancake-stack separation $a$ at
$a=[1-2]\lambda _{J}$, and rather wide range of the ratio
$\alpha$, $0.4 \lesssim \alpha\lesssim 0.65$. With decreasing $a$,
the crossing chain goes through intermediate strongly-deformed
configurations and smoothly transforms into a tilted chain via a
second-order phase transition. Another phase transition occurs at
very small densities of pancake vortices, $a\sim
[20-30]\lambda_J$, and only when $\alpha$ exceeds a certain
critical value $\sim 0.5$. In this case a small c-axis field
penetrates in the form of kinks. However, at very small
concentration of kinks, the kinked chains are replaced with
strongly deformed crossing chains via a first-order phase
transition. This transition is accompanied by a very large jump in
the pancake density.

\end{abstract}
\section{INTRODUCTION}

Layered superconductors have an amazingly rich phase diagram in
tilted magnetic field. Point (or ``pancake'') vortices  generated
by the c-axis component of the magnetic field \cite{pancakes} can
form a vast variety of lattice structures. Possible structures
include the kinked lattice,
\cite{IvlevMPL1991,BLK,Feinberg93,Kinkwalls}, tilted vortex chains
\cite{TiltedChains}, coexisting lattices with different
orientation\cite{coex-lat} and crossing lattices composed of
sublattices of Josephson vortices (JVs) and pancake-vortex
stacks\cite{BLK,CrossLatPRL99}. The main source of such richness
is the existence of two very different kinds of interactions
between the pancake vortices in different layers: magnetic and
Josephson interactions. The key parameter, which determines the
relative strength of these two interactions and plays a major role
in selecting the lattice structures, is the ratio of the two
fundamental lengths, the in-plane London penetration depth,
$\lambda \equiv\lambda_{ab}$, and Josephson length
$\lambda_{J}=\gamma s$, with $\gamma$ being the anisotropy
parameter and $s$ being the interlayer spacing, $\alpha=
\lambda/\lambda_{J}$. One can distinguish two limiting cases which
we refer to as ``extremely anisotropic'' case, $\alpha< 0.4$, and
``moderately anisotropic'' case $\alpha> 0.7$. Among known
atomically layered superconductors, only
Bi$_{2}$Sr$_{2}$CaCu$_{2}$O$_{x}$ (BSCCO) and related compounds
may belong to the ``extremely anisotropic'' family. Even in this
compound the parameter $\alpha$ is not smaller than $\sim0.25$ and
increases with temperature so that BSCCO typically becomes
``moderately anisotropic'' in the vicinity of the transition
temperature.
\begin{figure}[ptb]
\begin{center}
\includegraphics[width=3.4in]{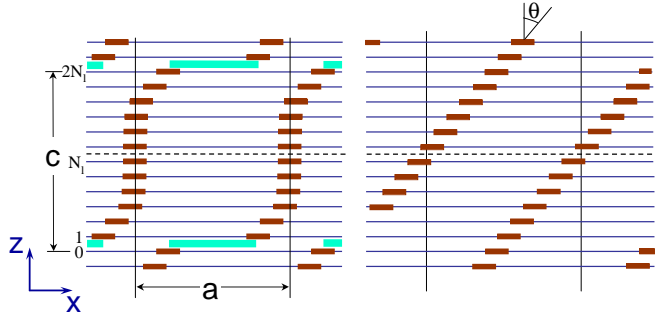}
\end{center}
\caption{Crossing (left) and tilted (right) vortex chains.}%
\label{Fig-CrossTiltChain}%
\end{figure}

At very small c-axis fields (up to 1-2 G) the pancake stacks in layered
superconductors within a wide range of anisotropies are arranged in
chains. An isolated chain is a two-dimensional array of pancake vortices
oriented perpendicular to the layers, see Fig.\ \ref{Fig-CrossTiltChain}.
At somewhat higher fields the chains are surrounded by the regions of
regular vortex lattice\cite{Bolle91}. The internal structure of an
isolated chain depends on the ratio $\alpha$ and it is relatively simple
in two limiting cases. At large $\alpha$ the chain is composed of tilted
pancake stacks (tilted chain, right picture in Fig.\
\ref{Fig-CrossTiltChain}) and at small $\alpha$ it consists of crossing
arrays of JVs and pancake stacks (crossing chains, left picture in Fig.\
\ref{Fig-CrossTiltChain}). In this paper we address the highly nontrivial
problem of how one structure transforms into another in the region of
intermediate $\alpha$. We analytically and numerically computed ground
state configurations in the isolated vortex chain and found two types of
phase transitions. The first phase transition typically takes place for
the intermediate separations between pancake stacks $a$, $a=[1-2]\lambda
_{J}$, and rather wide range of the ratio $\alpha$, $0.4 \lesssim
\alpha\lesssim0.65$. The ground state is given by the crossing chain in a
wide range of pancake separations $a$. Due to attraction between deformed
pancake stacks \cite{BuzdinPRL02}, the equilibrium separation can not
exceed some maximum value, which depends on the in-plane field and
$\alpha$ and it is typically of the order of several $\lambda_{J}$. With
decrease of the pancake separation $a$, the crossing chain becomes
strongly deformed and smoothly transforms into the modulated tilted
vortices. The modulation vanishes at a second-order phase transition at
which the system transforms into the tilted-chain state. This phase
transition provides possible interpretation of the recent
Lorentz-microscopy experiment \cite{MatsudaSci02} at which it was observed
that the pancake stacks located in the chains smear along the chain
direction while the pancake stacks outside chains remain well defined.
This smearing may indicate transition into the tilted-chain state.

Another phase transition occurs at very small densities of pancake
vortices, $a\sim 20-30\lambda_J$, and only when $\alpha$ exceeds a certain
critical value $\approx 0.5$ (exact criterion depends on the in-plane
magnetic field). In this case a small c-axis field penetrates in the form
of kinks. The kinked vortex lines forming tilted chains are composed of JV
pieces separated by kinks\cite{IvlevMPL1991,BLK,Feinberg93}. If the kink
energy is only slightly smaller then the energy per pancake in a straight
pancake stack then at very small concentration of kinks, typically at
$a\approx[20-30]\lambda_{J}$, the kinked chains are replaced with strongly
deformed crossing chains via a first-order phase transition. Due to the
opposite signs of interactions (kinks repel and deformed pancake stacks
attract each other) this transition is accompanied by a very large jump in
the pancake density. With further decrease of the pancake separation the
chain smoothly transforms back to the tilted chain as it was described in
the previous paragraph.

We constructed the chain phase diagrams for different ratios
$\alpha$. As follows from the above description, there are two types
of phase diagrams in the region of intermediate $\alpha$'s:
\begin{itemize}
\item In the range $0.4 \lesssim\alpha\lesssim0.5$ a small
$c$-axis field first penetrates in the form of pancake-stack
chains located on JVs. Due to attractive coupling between the
deformed stacks, their density jumps from zero to a finite value.
With further increase of the $c$ axis field the chain goes through
intermediate strongly deformed configurations and  smoothly
transforms into a tilted chain;%
\item In the range $0.5 \lesssim\alpha\lesssim0.65$ small $c$-axis
field first penetrates in the form of kinks creating kinked tilted
vortices. With increase of a $c$ axis field this structure is
replaced via the first-order phase transition with the
strongly-deformed crossing chains. This transition is accompanied
by a large jump of pancake density. Further evolution of the chain
structure is identical to the smaller-$\alpha$ scenario: the
structure smoothly transforms back into the tilted-chain state.
\end{itemize}
The exact transition between the two types of behavior depends
also on in-plane field. Larger in-plane field favors the first
scenario.

The second scenario provides a natural interpretation to recent
scanning-Hall-probe observations\cite{GrigNat01}. It was observed
that at very small concentration of pancakes the chains are
magnetically homogeneous and separate pancake stacks are not
resolved. When the external field exceeds some critical value of
the order of several Oersted, crystallites of the pancake stacks
are suddenly formed along the chain and the flux density in the
crystallites is much higher then the flux density in the
homogeneous chains. Our calculations provide consistent
interpretation for these observations. The magnetically
homogeneous chains are interpreted as kinked/tilted chains (such
interpretation has been proposed by Dodgson\cite{DodgsonPRB02})
and formation of crystallites can be attributed to the low-density
[kinked lines]-[crossing chains] first-order phase transition with
density jump.

The paper is organized as follows. In Section \ref{Sec-General} we
consider the chain energy functional. In Section \ref{Sec-Analyt}
we present analytical estimates  for the chain energy in the two
limiting cases, crossing and tilted chain, and discuss location of
transitional region between these states. In section
\ref{Sec-Attrac} we review attractive interaction between deformed
pancake stacks located on JVs. Section \ref{Sec-Numerical}
contains the results on numerical exploration of the the phase
diagram. We discuss two different phase transitions between the
tilted and crossing chains and two types of phase diagrams which
are realized in the region of intermediate parameter $\alpha$.

\section{CHAIN ENERGY FUNCTIONAL\label{Sec-General}}

In this paper we focus on the structure of an isolated vortex chain with
period $a$ in $x$ direction and period $c=Ns$ in $z$ direction where
$N=2N_l$ is the number of layers per vertical chain period (see Figure
\ref{Fig-CrossTiltChain}). This corresponds to the tilting angle $\theta$
of the magnetic induction with respect to the $c$ axis with $\nu\equiv
\tan\theta=a/c$. The pancake separation $a$ is determined by the c-axis
field $B_z$. Note that we consider the region of very small $B_z$ (0.1-5
G), where it is typically much smaller than external field. The vertical
period $c$ is fixed by the in-plane field $B_{x}$,
$c\approx\sqrt{2\Phi_{0}/(\sqrt{3}\gamma B_{x})}$. For BSCCO
($\gamma\sim500$) this period is approximately equal to $20$ layers at
$B_{x}\approx 50$G. The chains are separated by distance $c_y$,
$c_{y}=\Phi_{0}/cB_{x}\gg c$.

Our calculations are based on the Lawrence-Doniach free-energy
functional in the London approximation, $F_{\mathrm{LD}}$, (see,
e.g., Ref.\ \onlinecite{BLK}). The chain structure is mainly
determined by pancake displacements from aligned positions $u_n$
and regular phase distribution $\phi_{r,n}(x,y)$. We consider the
case $c\ll\lambda$ and in-plane distances much smaller than
$\lambda_{c}$. In this situation the general Lawrence-Doniach
energy, $F_{LD}$, can be significantly simplified using several
approximations: (i)we neglect screening of regular phase and
$z$-axis vector-potential; (ii)we subtract the energy of straight
pancake stacks, $(B_{x}/\Phi_{0})\varepsilon _{\mathrm{PS}}$,
allowing us to eliminate logarithmically diverging pancake-core
contributions; (iii)we drop the trivial magnetic energy term
$B_{x}^{2}/8\pi$ which plays no role in selection between
different chain phases. We will use the chain energy per unit area
lattice, $E\equiv c_y[F_{\mathrm{LD}}/V-B_{x}^{2}/(8\pi)-(B_{z}%
/\Phi_{0})\varepsilon_{\mathrm{PS}}]$ with $V$ being the total
system volume. With the above assumptions the chain energy
functional can be represented as
\begin{align}
E & =\frac{1}{c}\sum_{n=1}^{N}\int\limits_{0}^{a}\frac{dx}{a}\int
\limits_{-c_{y}/2}^{c_{y}/2}dy \left[  \frac{J}{2}\left(
\nabla\phi_{r,n}\right)  ^{2}\right.
\nonumber\\
&\left. +E_{J}\left(  1-\cos\left( \nabla_{n}\left( \phi
_{r,n}+\phi_{v,n}\right)  -\frac{2\pi s}{\Phi_{0}}B_{x}y\right)
\right)\right]  \nonumber\\
&  +\frac{1}{2}\frac{N}{N_{tot}}\sum_{n\neq
m}U_{Mr}(u_{n}-u_{m},n-m)\label{ChainEn}
\end{align}
where the phase stiffness, $J$, and the Josephson coupling energy,
$E_J$, are given by
\begin{equation}
J\equiv\frac{s\varepsilon_{0}}{\pi},\;E_{J}\equiv\frac
{\varepsilon_{0}}{\pi s\gamma^{2}}, \;\text{with }\varepsilon_{0}
\equiv\frac{\Phi_{0}^{2}}{( 4\pi\lambda)^{2}},
\end{equation}
$\lambda \equiv\lambda_{ab}$ and $\lambda_{c}$ are the components of the
London penetration depth and $\gamma=\lambda_{c}/\lambda_{ab}$. The ratio
of the two energy scales determines the most important length scale of the
problem, the Josephson length, $\lambda_{J}=\gamma s=\sqrt{J/E_{J}}$.
$\phi_{v,n}\left( \mathbf{r}\right) $ is the vortex phase variation
induced by displacement of pancake rows, $u_{n}$, from the ideally aligned
positions
\begin{equation}
\varphi_{v,n}(x,y;u_{n})=\arctan\frac{\tan\left(
\pi(x+u_{n})/a\right) }{\tanh\left(  \pi y/a\right)
}-\arctan\frac{\tan\left(  \pi x/a\right)
}{\tanh\left(  \pi y/a\right)  };\label{VortRowPhase}%
\end{equation}
the discrete gradient $\nabla_{n}\phi_{n}$ is defined as
$\nabla_{n}\phi _{n}\equiv\phi_{n+1}-\phi_{n}$.
$U_{Mr}(u_{n}-u_{m},n-m)$ is the interaction energy between the
pancake rows per unit length, computed with respect to straight
stacks
\[
U_{Mr}(u,n)\equiv\frac{1}{a}\sum_{m}\left[
U_{M}(ma+u,n)-U_{M}(ma,n)\right] ,
\]
where $U_{M}(\mathbf{R},n)$ is the magnetic
interaction between pancakes\cite{pancakes}%
\begin{align}
U_{M}(\mathbf{R},n)  &  \approx2\pi J\left[  \ln\frac{L}{R}\left[
\delta _{n}-\frac{s}{2\lambda}\exp\left(
-\frac{s|n|}{\lambda}\right)  \right] +\frac{s}{4\lambda}u\left(
\frac{r}{\lambda},\frac{s|n|}{\lambda}\right)
\right] \label{MagInter}\\
u\left(  r,z\right)   &  \equiv\exp(-z)E_{1}\left(  r-z\right)
+\exp (z)E_{1}\left(  r+z\right)  ,\nonumber
\end{align}
$E_{1}(u)=\int_{u}^{\infty}\left(  \exp(-v)/v\right)  dv$ is the
integral exponent, $r\equiv\sqrt {R^{2}+(ns)^{2}}$, and $L$ is a
cutoff length; $N_{tot}$ is the total number of layers.

The energy (\ref{ChainEn}) contains contribution coming from a
long-range suppression of the Josephson energy accumulated from
distances $c\ll y\ll c_{y}$, that is identical in all chain
phases, $E_{\text{\textsc{J-LR}}}=E_{J}\pi^{2}c_{y}/(6N^2s)$. It
is convenient to separate this term and define the local chain
energy, $E^{loc}$,
\begin{equation}
E^{loc}\equiv E-E_{\text{\textsc{J-LR}}}.
\label{Local-En}%
\end{equation}
This part of energy weakly depends on $c_{y}$ and does not diverge
for $c_{y}\rightarrow\infty$.

We numerically minimized the energy functional (\ref{ChainEn})
with respect to $u_n$ and $\phi_{r,n}(x,y)$ within unit cell,
$0<x<a$, $0<y<c_y/2$, and $0<n<N_l=N/2$ using appropriate symmetry
and periodic boundary conditions. In calculation of magnetic
coupling energy one has to take into account periodic conditions
for pancake displacements, $u_{n+N}=u_{n}$. In addition, if one
selects $z$ axis origin at the center of the JV then symmetry also
requires $u_{-n}=-u_{n}$. Two simple limiting cases in Fig.\
\ref{Fig-CrossTiltChain} correspond to (i) $u_{n}\ll a$ for the
crossing chain and (ii) $u_{n}=-a(1-(n-1/2)/N_{l})/2$ for the
tilted chain.

\section{CHAIN ENERGIES IN LIMITING CASES \label{Sec-Analyt}}

Analytic estimates for energy contributions are possible in two limiting
cases of weakly deformed crossing chain and tilted chain shown at Fig.\
\ref{Fig-CrossTiltChain}. In this section we summarize results for chain
energies in different limits. Detailed derivations will be published
elsewhere.

Local energy of crossing chain per unit area is given by
\begin{equation}
E_{CL}^{loc}=\frac{\varepsilon_{0}}{a}\left(
\frac{\nu}{\gamma}\left(  \ln N-0.41\right)
-\frac{8\alpha^{2}}{\ln(3.5/\alpha)N}\right)  ,
\label{EnCrossChain_loc}%
\end{equation}
Here the first term represents the local JV energy, $E^{loc}_{JV}$, and
the second term is the contribution from the crossing energies of JVs and
pancake stacks\cite{CrossLatPRL99}.

The local tilted-chain energy per unit area in the limit
$\nu\equiv \tan \theta \ll \gamma$ is given by
\begin{equation}
E_{TV}^{loc}=\frac{\varepsilon_{0}}{a}\left[  U\left( \frac{a}{2\pi\lambda
}\right)  +\frac{\nu^{2}}{2\gamma^{2}}\left( \ln N-0.95\right)  \right],
\label{E_TVsm_local}%
\end{equation}
\[\text{where  }
U\left(  x\right)  =\sum_{m=1}^{\infty}\left(  \frac{1}{m}-\frac{1}%
{\sqrt{m^{2}+x^{2}}}\right)  =
\genfrac{\{}{.}{0pt}{}{\zeta(3)x^{2}/2,\;x\lesssim0.5}{\frac{1}{2x}-\ln
\frac{2}{x}+\gamma_{E},\;x\gtrsim1},
\]
with $\gamma_{E}\approx0.5772$ being the Euler constant. In Eq.\
(\ref{E_TVsm_local}) the first term represents the loss of the
magnetic coupling energy in the tilted chain and the second term
represents the Josephson energy loss.

In the region of kinked lines $\nu\gg\gamma$ (but $\nu\ll N\gamma/2\pi$),
the local chain energy is given by
\begin{equation}
E^{loc}_{TV}\approx E^{loc}_{JV}+\frac{\varepsilon_{0}}{a}\left[ \ln
\frac{0.44}{\alpha}+\frac{\gamma}{2\nu}\left( \ln\left( \frac{\gamma
N}{\nu}\right)-2.454\right) \!\right].
\label{E_TVresult_lrg}%
\end{equation}
In this equation $E^{loc}_{JV}$ is the local JV-lattice energy, the first
term in square brackets gives the single-kink energy and the third term
gives the kink interaction energy.

To find out whether the crossing or tilted chain is realized for given
values of parameters $a$, $c$, and $\alpha$, we have to compare the
energies of these states. Comparison of Eqs.\ (\ref{EnCrossChain_loc}) and
(\ref{E_TVsm_local}) gives transitional pancake separation $a_c$ which
decreases with $N$ and increases with $\alpha$ \cite{RioProc}. Naively,
one may think that intersection of the energy curves for two states would
correspond to a first-order phase transition between these states.
However, as we will see from numerical simulations, in the region
$\nu=\tan\theta<\gamma$ another scenario is realized. Typically, strongly
deformed intermediate chain configurations develop in the transitional
region providing a smooth transformation between the two limiting
configurations. Therefore, a simple energy comparison gives only an
approximate location of the transitional region separating the two
configurations.

In the region $1<\nu/\gamma<N/2\pi$ and $a>2\pi\lambda$ comparison
of the energies (\ref{EnCrossChain_loc}) and (\ref{E_TVresult_lrg})
gives the following equation
\begin{equation}
\ln\frac{1}{\alpha}-0.81+\frac{\gamma}{2\nu}\left(  \ln\left(
\frac{\gamma N}{\nu}\right)  -2.454\right)
+\frac{8\alpha^{2}}{\ln(3.5/\alpha
)N}=0\! \label{Tran_Crit_Lrg}%
\end{equation}
This equation has a solution only in the kink penetration regime,
$\ln\left( 1/\alpha\right) -0.81<0$, where the kink energy is only
slightly smaller that the energy per pancake of a straight
pancake-vortex stack. In contrast to the case $\nu \ll \gamma $,
this equation does correspond to a very strong first-order phase
transition.

\section{ATTRACTION BETWEEN DEFORMED PANCAKE STACKS.
MAXIMUM EQUILIBRIUM SEPARATION \label{Sec-Attrac}}

A peculiar property of the crossing chain is an attractive
interaction between the deformed pancake stacks at large
distances\cite{BuzdinPRL02}. As a consequence, when the magnetic
field is tilted from the direction of layers the density of
pancake stacks located on JVs jumps from zero to a finite value.
This means the existence of a maximum equilibrium separation
$a_{m}$ between pancake stacks, i.e., chains with $a>a_{m}$ are
not realized in equilibrium. Note that the tilted vortices also
attract each other within some range of angles and distances
\cite{TiltedChains} meaning that tilted chains also have this
property in some range of parameters.

Simple analytical formula for the attraction energy between the
deformed pancake stacks can be derived for very anisotropic
superconductors $\lambda_{J}\gg\lambda$ in the range $\lambda\ll
R\ll\lambda_{J}$ \cite{BuzdinPRL02}. In this limit short-range
pancake displacements $u_{n}$ from the aligned positions in two
neighboring stacks produce dipole-like contribution to interaction
energy per unit length between these stacks $ \delta U_{i}(R)
=-2\varepsilon_{0}\left\langle u^{2}\right\rangle /R^{2}. $ This
term has to be combined with the usual repulsive interaction between
straight stacks $U_{i0}(R)=2\varepsilon_{0}\mathrm{K}_{0}(R/\lambda
)\approx2\varepsilon_{0}\sqrt{\pi\lambda/2R}\exp\left(
-R/\lambda\right)  $. Minimum of the total interaction energy,
$U_{i0}(R)+\delta U_{i}(R)$ gives an estimate for the maximum
equilibrium separation $a_{m}$ \cite{BuzdinPRL02},
$a_{m}=\lambda\ln(C\lambda^{2}/\left\langle u^{2}\right\rangle )$
and this result is valid until $a_{m}<\lambda_{J}$. Because in BSCCO
$\lambda_{J}$ is only $2-3$ times larger than $\lambda$, this simple
formula is not practical for this compound. We will see that $a_{m}$
in BSCCO is usually larger than $\lambda_{J}$. In general, the
maximum equilibrium separation $a_{m}$ is determined by the minimum
of  the pancake part of energy per one stack $U(a)=a\left(
E(a)-E(\infty)\right)$. When the main contribution to the total
interaction energy is coming from the nearest-neighbor interaction,
$a_{m}$ coincides with the position of the minimum in the pair
interaction potential.

\section{NUMERICAL EXPLORATION OF CHAIN STRUCTURES \label{Sec-Numerical}}

We explored the chain structures by numerically minimizing
(\ref{ChainEn}) with respect to the pancake row displacement
$u_{n}$ and regular phase distribution $\phi_{r,n}$ for different
values of the parameters $a$, $N=2N_{l}$, and $\alpha$.  In the
following sections we review the results of these calculations.

\subsection{Phase Transition From Crossing To Tilted Chains With
Decreasing Pancake Separation \label{Sec-PhaseDiag_sm}}

\begin{figure}[ptb]
\begin{center}
\includegraphics[width=4.5in]{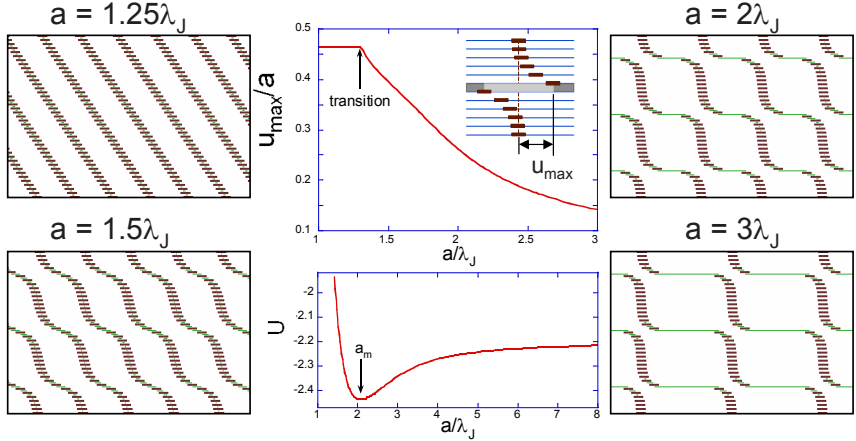}
\end{center}
\caption{\emph{Upper plot} shows the dependence of the maximum
displacement(defined in the inset) on pancake separation $a$ for
$N=14$ and $\alpha=0.4$. \emph{Lower plot} shows dependence of the
pancake energy per unit cell $U$ (in units of phase stiffness $J$)
on pancake separation. Its minimum determines the maximum
equilibrium separation $a_m$. Representative chain structures are
shown for several values of $a$(boxes show positions of the
pancake vortices and horizontal lines mark JVs). One can see that
the system evolves from weakly deformed chain ($a=3\lambda_J$) via
strongly deformed chain ($\alpha=2\lambda_J$) to modulated tilted
chain ($\alpha=1.5\lambda_J$). The last structure transforms via a
2nd-order phase transition at $\alpha=1.3\lambda_J$ into tilted
straight vortices.}
\label{Fig-Nl7al0_4}%
\end{figure}

We studied the evolution of chain structures with decreasing
pancake separation $a$ at fixed $\alpha$ and $N$.  For small
values of $\alpha$ we found that the chain structure smoothly
evolves with decreasing $a$ from crossing to tilted configuration.
An example of such evolution is presented in Fig.\
\ref{Fig-Nl7al0_4} for $N=14$ and $\alpha=0.4$. The upper plot
shows the dependence of the maximum pancake displacement from the
straight-stack position $u_{max}/a$ (defined in the inset) on the
pancake separation $a$. We will use this parameter to characterize
the chain structure throughout the paper. It changes from zero for
straight stacks to $(1-1/N)/2$ for tilted chains. The lower plot
shows the $a$ dependence of the reduced pancake energy per unit
cell, $U\equiv ca(E(a)-E(\infty))/J$, which is used to determine
the maximum equilibrium separation $a_m$. At large $a$ a weakly
deformed crossing configuration is realized (see structure at
$a=3\lambda_J$). With decrease of $a$ the chain evolves into
strongly corrugated configurations such as configuration for
$a=2\lambda_J$ in Fig.\ \ref{Fig-Nl7al0_4}. With further decrease
of $a$ this structure smoothly transforms into modulated tilted
lines (see structure for $a=1.5\lambda_J$). Finally, the last
structure transforms via a second-order phase transition into the
straight tilted lines. For parameters used in Fig.\
\ref{Fig-Nl7al0_4} this occurs at $a=1.3\lambda_J$. The plateau in
the dependence $u_{max}(a)/a$ below this value of $a$ corresponds
the maximum possible relative displacement $(1-1/N)/2\approx
0.4643$ in the tilted chain.

Comparison of numerical and analytical calculations shows that the
analytical estimates (\ref{EnCrossChain_loc}) and
(\ref{E_TVsm_local}) accurately reproduce numerical results for
the weakly deformed crossing chain and for the tilted chain.
However, the numerical study predicts intermediate configurations
with energies smaller than the energies of the both limiting
configurations. As a result, a naively expected first-order phase
transition is replaced by a continuous transition occurring at
significantly smaller $a$. The transitional region just marks the
location of the intermediate strongly deformed chain
configurations. The observed continuous phase transition is
related to the instability of isolated tilted vortices in
anisotropic
superconductors\cite{ThompsMoorePRB97,BenkraoudaPRB96}.

\begin{figure}[ptb]
\begin{center}
\includegraphics[width=3.in]{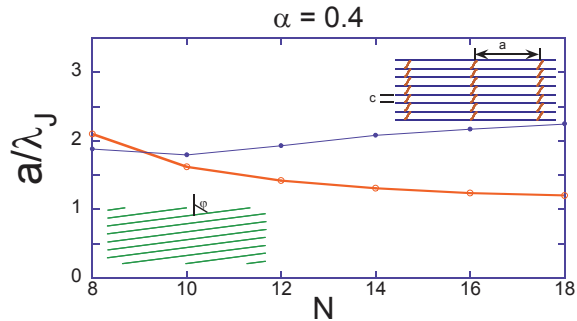}
\end{center}
\caption{Chain phase diagrams in the plane $a/\lambda_{J}-N$ for
the ratio $\alpha=0.4$. Thick line with open circles indicates
phase transition into the tilted chain. Thin line shows
the maximum equilibrium separation $a_{m}$.}%
\label{Fig-PhasDiagAl04}%
\end{figure}
In Fig.\ \ref{Fig-PhasDiagAl04} we present the chain phase
diagrams in the $a$-$N$-plane for $\alpha=0.4$. The thick line
shows the phase transition into the tilted-chain state. One can
see that at larger $N$ the transition takes place at smaller $a$.
With increase of $\alpha$ this line moves higher. Dotted lines
show locations of the maximum equilibrium separation $a_{m}$
discussed in Sec.\ \ref{Sec-Attrac}. We see that $a_{m}(N)$ line
crosses the transition line meaning that at small $N$ $a_{m}$
falls into the tilted-chain region and at large $N$ it falls into
the crossing-chain region. The obtained phase diagram implies that
at small tilting angle of the field (corresponding to small $a$)
the tilted chains have lower energy than the crossing chains. This
is similar to the situation at higher fields, in the dense
lattice, where the crossing-lattices state also is expected to
transform into the simple tilted lattice at small tilting angle of
the field\cite{CrossLatPRL99}.

\subsection{Reentrant Transition To Kinked/Tilted Lines At Small
Concentration Of Pancakes At $\alpha \gtrsim 0.5 $}

At higher values of the ratio $\alpha$ a new qualitative feature appears
in the phase diagram. When $\alpha$ exceeds the critical value, the $c$
axis field initially penetrates in the form of kinks forming kinked vortex
lines (lock-in transition\cite{IvlevMPL1991,Feinberg93,BLK}). The critical
value of $\alpha$ is determined by combination of numerical constants in
the pancake-stack and kink energies. At present,  our best estimate for
this constant is $\alpha_c\approx 0.44$. The critical value of $\alpha$
increases with decrease of $N$, due to the increasing relative
contribution of the crossing energies in the energy of crossing chain. A
very peculiar behavior is expected when $\alpha$ only slightly exceeds
$\alpha_{c}$. The competing chain states have very different interactions:
deformed stacks attract and kinks repel each other. Moreover, at the same
value of the c-axis magnetic induction, $B_{z}$, the kinks are much closer
than the stacks and the absolute value of the kink interaction energy is
much larger than the interaction energy between deformed stacks. As a
consequence, with increase of $B_{z}$ the total energy of the kinked lines
rapidly exceeds the total energy of the crossing chain and the system
experiences a first-order phase transition into the crossing-chain state.
Due to the attractive interaction between the pancake stacks, the
pancake/kink separation at which the energy curves cross does not give the
equilibrium separation for the crossing chain and the stack separation
jumps at the transition to a value slightly smaller than the maximum
equilibrium separation $a_{m}$. This means that the phase transition is
accompanied by jumps of the pancake density and magnetic induction,
$B_{z}$.

\begin{figure}[ptb]
\begin{center}
\includegraphics[width=4.3in]{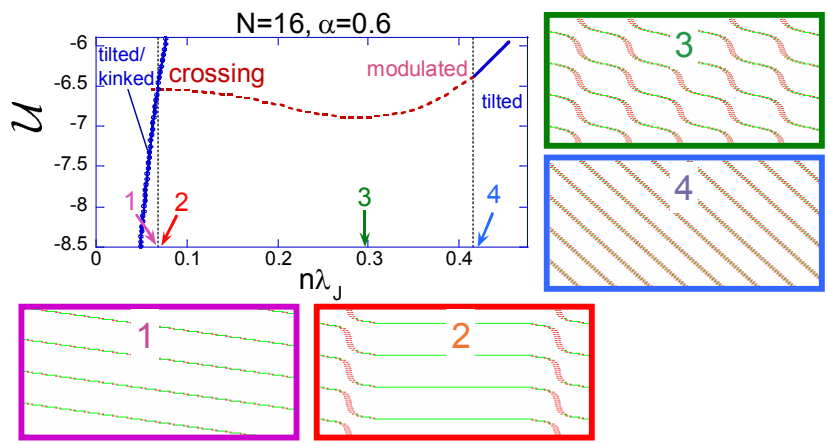}
\end{center}
\caption{The dependence of the pancake energy per chain unit cell
on pancake density $n=1/a$ for $N=16$ and $\alpha=0.6$. The two
branches at low $n$ correspond to two different starting states,
kinked lines (1) and crossing chain (2). The kinked lines  have
lower energy at very small $n$, at $n<0.068$. Crossing chain
smoothly transforms back into tilted chain with increase of $n$.
The transformation is completed at the second-order phase
transition point near $n\lambda_{J}=0.4$. Chain configurations are
shown at four marked points. }
\label{Fig-adEN16Al0_6}%
\end{figure}%
This behavior was confirmed by numerical calculations. Figure
\ref{Fig-adEN16Al0_6} shows a plot of the dependence of the energy per
chain unit cell $\mathcal{U}$ on the pancake density $n=1/a$ for $N=16$
and $\alpha=0.6$. This dependence has two branches, corresponding to two
different starting states at small $n$, crossing chain and kinked lines,
and the kinked vortex lines have smaller energy at smaller $n$. The
branches cross at $n=0.068$ marking the first-order transition. The
variations of $\mathcal{U}$ at small $n$ occur due to the interaction
energy and one can see that the kink interaction energy is much larger
than the crossing-chain interaction energy. With further increase of $n$,
the crossing chain smoothly transforms into the tilted chain following the
scenario described in the previous Section. The second-order phase
transition for these parameters takes place at $n\approx 0.4$
corresponding to $a\approx 2.5\lambda_{J}$. The latter value is somewhat
smaller than the maximum equilibrium separation
$a_{m}\approx 3.44\lambda_{J}$.%
\begin{figure}[ptbptb]
\begin{center}
\includegraphics[width=3.9in]{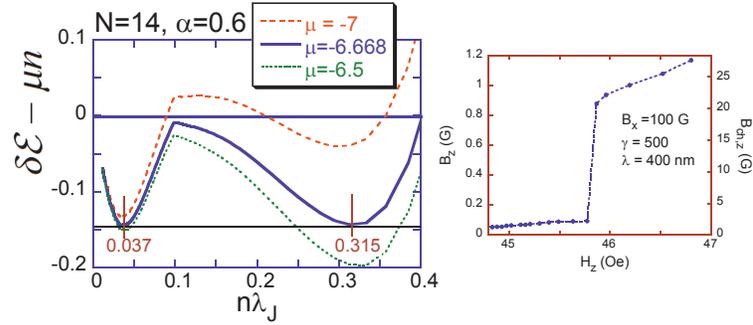}
\end{center}
\caption{\emph{Left plot} shows the density dependence of the pancake part
of the thermodynamic potential per unit area, $\delta\mathcal{E}-\mu n$
(in units of $\epsilon _{0}/(\pi\lambda_{J} N)$) at different values of
chemical potential $\mu$ corresponding to different external fields for
$N=14$ and $\alpha=0.6$. Kinks in the curves separate regions of
tilted/kinked lines (low $n$) and crossing chains (high $n$). The
equilibrium density is given by the global minimum of this energy. One can
see that at $\mu \approx-6.668$ the system experiences a first-order phase
transition with very large density jump. \emph{Right plot} shows the
corresponding dependencies of the average magnetic induction, $B_z$, (left
axis) and the maximum induction in the middle of the chain, $B_{ch,z}$,
(right axis) vs the magnetic field strength, $H_z$.}
\label{Fig-dE-munN16Al0_6}%
\end{figure}%
\begin{figure}[ptb]
\begin{center}
\includegraphics[width=4.in]{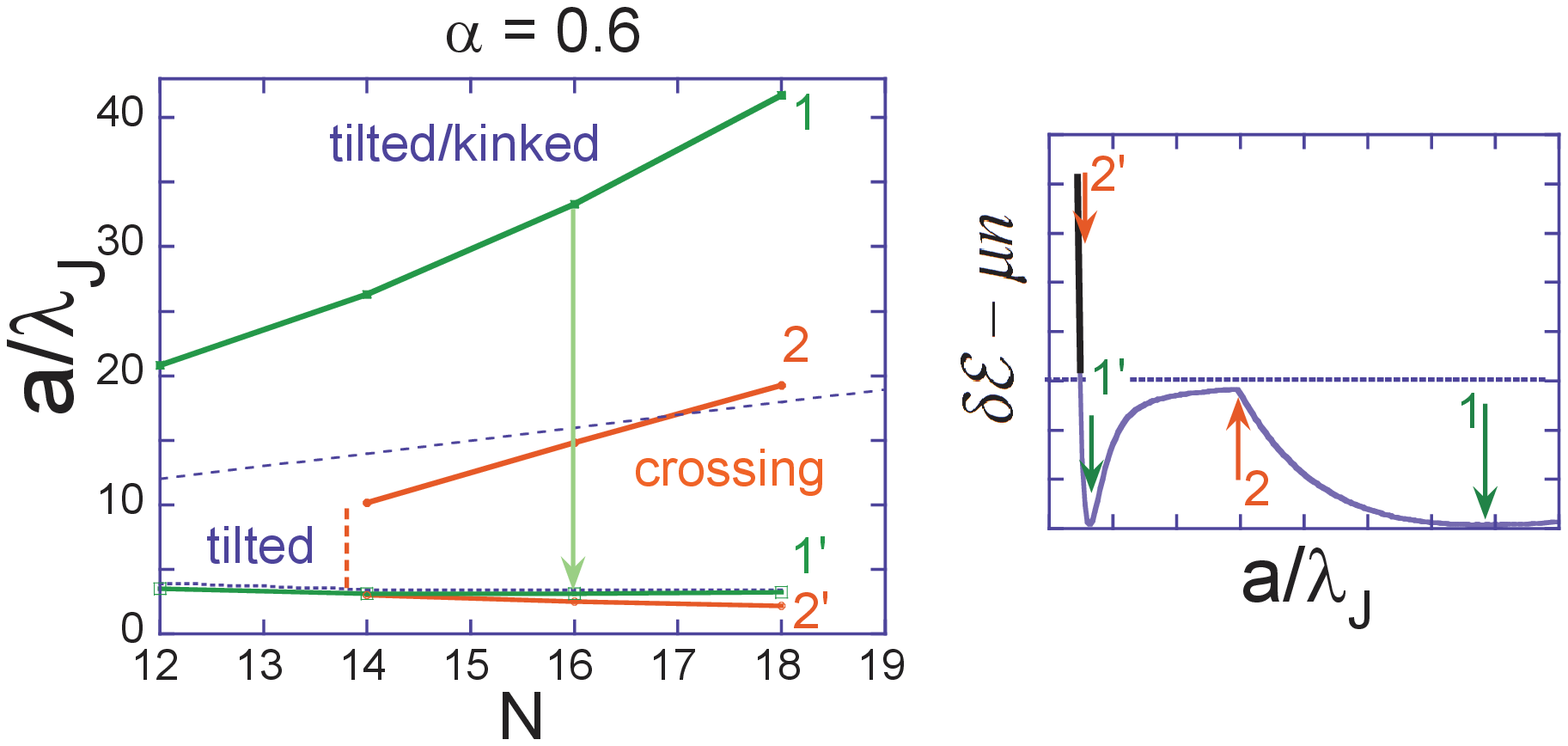}
\end{center}
\caption{\emph{The left panel} shows the phase diagram in the
$N$-$a$ plane for $\alpha=0.6$. \emph{The right panel} illustrates
definitions of the phase lines using the plot of the thermodynamic
potential $\delta\mathcal{E}-\mu n$ vs $a/\lambda_{J}$ at the
transition point. The lines $1$ and $1^{\prime}$ correspond to the
two limiting pancake separations at the transition point between
which the jump occurs. The line 2 indicates crossing of the energy
curves for the kinked and crossing chain. The line $2^{\prime}$
shows position of the continuous transition into the tilted chain.
Dotted line slightly above $1^{\prime}$-line shows position of the
maximum equilibrium separation $a_{m}$. We also show by the dashed
line the crossover line $a/\lambda_{J}=N$ above which well-defined
kinks appear.}
\label{Fig-PhaseDiagAl0_6}%
\end{figure}%

The pancake density in the chain can not be directly fixed in
experiment. Instead, the magnetic field strength, $H_z$, fixes the
chemical potential $\mu_{H}$, $\mu_{H}=\Phi_{0} H_z/(4\pi)$, and
the equilibrium density is determined by the global minimum of the
thermodynamic potential $G(n)=E(n)-\mu_{H} n$. To find the density
evolution with increasing the chemical potential, we plot in Fig.\
\ref{Fig-dE-munN16Al0_6}(left) the density dependencies of the
reduced thermodynamic potential, $\delta\mathcal{E} -\mu n$ for
different $\mu$ and representative parameters $N=14$ and
$\alpha=0.6$. As the energy of isolated stacks is subtracted in
$\delta\mathcal{E}$, the dimensionless chemical potential is
shifted with respect to its bare value and it is related to the
magnetic field strength
as %
$\mu=N\Phi_{0}(H_z-H_{c1})/(4\epsilon_0)$, %
where $H_{c1}$ is the lower critical field for $\mathbf{H}\| c$. We find
that for selected parameters the transition takes place at
$\mu=\mu_{t}=-6.668$. At $\mu<\mu_{t}$ the global minimum falls into the
region of kinked lines and at $\mu>\mu_{t}$ it jumps into the region of
crossing chain. Note that the density value, at which the energy curves
cross (kinks in the lines in Fig.\ \ref{Fig-dE-munN16Al0_6}), is always
larger than the lower density from which the jump to the high-density
state takes place. At the transition, the density jumps almost ten times,
from $0.037/\lambda _{J}$ to $0.315/\lambda_{J}$. The right plot in Fig.\
\ref{Fig-dE-munN16Al0_6} shows the corresponding dependencies of the
average magnetic induction, $B_z=\Phi_0n/c_y$, (left axis) and the maximum
induction in the middle of the chain, $B_{ch,z}=\Phi_0n/\lambda$, (right
axis) vs the magnetic field strength, $H_z$. One can see that the jump in
$B_z$ occurs from $\sim 0.1$G to $\sim 1$G.

The numerically obtained phase diagram in the $N$-$a$ plane for
$\alpha=0.6$ is shown in the left panel of Fig.\
\ref{Fig-PhaseDiagAl0_6}. The plot of the thermodynamic potential
at the transition point in the left panel illustrates definitions
of different lines in the phase diagram. The lines $1$ and
$1^{\prime}$ show the limiting pancake separations at the
first-order transition between which the jump takes place. When
the chemical potential is fixed by external conditions the area
between these lines is bypassed in equilibrium. The line $2$
inside this region marks the crossing of the energy curves for the
two states. At large $N$ the jump takes place from the
kinked-lines state into the strongly corrugated configuration.
This configuration transforms into the tilted chain with further
decrease of $a$ via continuous transition shown by the line
$2^{\prime}$. Below $N=14$ only tilted chains realize, but the
density jump still exists. The upper separation grows
approximately proportional to $N$ while the lower separation
slowly decreases with $N$ and lies slightly below the maximum
equilibrium separation $a_{m}$ shown by a dotted line. This means
that the relative density jump increases with $N$. This type of
phase diagram exists within a finite range of $\alpha$, roughly
$0.5\lesssim \alpha \lesssim 0.7$. At larger $\alpha$'s only
tilted chains exist.

\section{CONCLUSIONS}

In conclusion, we investigated the phase diagram of an isolated
vortex chain in layered superconductors. In the region where
Josephson and magnetic coupling are approximately equal, we found
a very rich behavior. The crossing chains typically transform into
the tilted chains with decreasing pancake separation via formation
of intermediate strongly deformed configurations and continuous
phase transition. When the relative strength of the Josephson
coupling exceeds some typical value, the phase diagram becomes
reentrant. At very small c-axis field, tilted chains are realized
in which the vortex lines have kinked structure. With increasing
the c-axis field these low-density tilted chains transform via a
first-order phase transition into the strongly-deformed crossing
chains. This transition is accompanied by a large jump of
pancake-vortex density. With further field increase these crossing
chains transform back into the tilted chains via a second-order
transition.

\section{ACKNOWLEDGEMENTS}

I would like to thank M.\ Dodgson, A.\ Grigorenko, S.\ Bending,
and V.\ Vlasko-Vlasov for useful discussions and the referee A for
improving presentation. I also gratefully acknowledge use of
``Jazz'', a 350-node computing cluster operated by the Mathematics
and Computer Science Division at Argonne National Laboratory as
part of its Laboratory Computing Resource Center. This work was
supported by the U.\ S.\ DOE, Office of Science, under contract \#
W-31-109-ENG-38.

\end{document}